\documentclass[a4paper]{article}
\usepackage{ASVspoof}
\usepackage{epsfig,amssymb,amsmath}
\usepackage{xcolor}
\usepackage{hyperref}
\usepackage{multirow}
\ninept

\title{Spoofing-Robust Speaker Verification Using Parallel Embedding Fusion: BTU Speech Group's Approach for ASVspoof5 Challenge}

\makeatletter
\def\name#1{\gdef\@name{#1\\}}
\makeatother

\name{{\em Oğuzhan Kurnaz\textsuperscript{1,2}, Selim Can Demirtaş\textsuperscript{1}, Aykut Büker\textsuperscript{1},}\\
      {\em Jagabandhu Mishra\textsuperscript{2}, Cemal Hanilçi\textsuperscript{1}}}

\address{\textsuperscript{1}Department of Electrical and Electronic Engineering,
Bursa Technical University, Bursa, Turkey \\
\textsuperscript{2}School of Computing,
University of Eastern Finland, Joensuu, Finland \\
{\small \tt \hspace{-0.8cm}\{oguzhan.kurnaz, selim.demirtas, aykut.buker, cemal.hanilci\}@btu.edu.tr, jagabandhu.mishra@uef.fi }}

\begin{document}
\maketitle

\begin{abstract}
This paper introduces the parallel network-based spoofing-aware speaker verification (SASV) system developed by BTU Speech Group for the ASVspoof5 Challenge. The SASV system integrates ASV and CM systems to enhance security against spoofing attacks. Our approach employs score and embedding fusion from ASV models (ECAPA-TDNN, WavLM) and CM models (AASIST). The fused embeddings are processed using a simple DNN structure, optimizing model performance with a combination of recently proposed a-DCF and BCE losses. We introduce a novel parallel network structure where two identical DNNs, fed with different inputs, independently process embeddings and produce SASV scores. The final SASV probability is derived by averaging these scores, enhancing robustness and accuracy. Experimental results demonstrate that the proposed parallel DNN structure outperforms traditional single DNN methods, offering a more reliable and secure speaker verification system against spoofing attacks. 
\end{abstract}

\section{Introduction}


\emph{Automatic speaker verification} (ASV) is a biometric technology used to authenticate a speaker's identity based on his/her voice characteristics \cite{reynolds94_asriv}. ASV systems are widely employed in various sectors, including banking for secure transactions, telecommunications for user authentication, and access control systems to grant entry to authorized individuals. By analyzing unique vocal features, ASV systems provide an efficient and user-friendly method for identity verification. However, with advancements in technology, spoofing attacks, where an imposter mimics a legitimate user's voice, have also evolved, posing significant threats to ASV systems. These attacks can exploit the vulnerabilities of ASV systems, leading to unauthorized access and security breaches. To counteract these threats, \emph{spoofing countermeasure} (CM) techniques \cite{kinnunen2017, Todisco2019} have been developed. CM systems are designed to detect and prevent spoofing attacks by distinguishing between bonafide and spoof voice samples, thereby enhancing the security of ASV systems.

In recent years, the integration of ASV and CM systems, known as \emph{spoofing-aware speaker verification} (SASV) \cite{Shim2022}, has gained attention. This combined approach aims to leverage the strengths of both ASV and CM to create a more robust and secure verification system. The primary focus of current research is to develop effective methods for integrating these two systems, ensuring that ASV systems are not only accurate in verifying legitimate users but also resilient against sophisticated spoofing attacks \cite{zhang2022backend, Liu2024}. This integration represents a significant advancement in the field, promising improved security and reliability for voice-based authentication systems.

Firstly, to develop spoofing-aware speaker verification, the \emph{SASV2022 challenge} has been announced \cite{Shim2022}. All participants aimed to design a spoofing-aware speaker verification system. The SASV2022 challenge introduced baseline systems that utilized both score-level and embedding-level fusion methods. The Baseline 1 (B1) employed a straightforward score-sum fusion strategy, combining ASV cosine similarity scores and CM output scores. This method is widely used and requires neither training nor fine-tuning. The Baseline 2 (B2) leveraged a more complex DNN-based back-end fusion strategy, which combined speaker and spoofing embeddings using a neural network. This approach aimed to create a joint representation space to enhance system performance. In \cite{Zhang2022}, a probabilistic fusion framework was introduced to integrate CM and ASV scores. This framework included fusion techniques for both direct inference and fine-tuning of SASV scores. An integration network was proposed in \cite{Martin-Donas2022}, where concatenated speaker verification and spoofing embeddings were input into a DNN to generate a low-dimensional spoofing embedding. The CM scores were computed as the cosine similarity between this low-dimensional spoofing embedding and the trained network parameters. Subsequently, the CM and ASV scores were linearly combined to derive the SASV scores. Various score-level fusion methods and cascade systems were suggested for the SASV challenge in \cite{Wang2022b}. Two back-end systems have been proposed in \cite{Heo2022a}: the Multi-Layer Perceptron Score Fusion Model (MSFM) and the Integrated Embedding Projector (IEP). The MSFM, a score fusion back-end system, derives the SASV score by utilizing both ASV and CM scores and embeddings. In contrast, the IEP combines ASV and CM embeddings into a unified SASV embedding and calculates the final SASV score based on cosine similarity.

Recently, the fifth edition of the biennial ASVspoof5 challenge \cite{Wang2024_ASVspoof5} was announced. One of the subtasks of this competition is the spoofing-aware speaker verification (SASV) (Track - 2) task. In this paper, we describe the several systems developed by Bursa Technical University (BTU) Speech Group for the SASV section of the competition. We utilized ECAPA-TDNN \cite{Desplanques_2020}, WavLM \cite{Chen2021WavLM}, and AASIST \cite{jung2022aasist} models to obtain scores and embeddings. By employing score fusion methods, we fused these scores and calculated a-DCF \cite{shim2024adcf} scores, the main metric announced for the competition. Furthermore, using the embeddings obtained from these models, we performed embedding fusion and trained the baseline2 DNN model proposed in the SASV2022 challenge. However, training a single DNN model for the SASV task using concatenated ASV and CM embeddings may be insufficient since stacking different types of embeddings (ASV and CM embeddings) yields heterogeneous input features. A single SASV model may struggle to effectively learn from such diverse input features. Hence, in this paper, we propose using two identical parallel DNN models with different inputs for the SASV task. With this approach, each DNN model can focus on capturing the unique patterns and relationships within its input features. The SASV scores produced by each model are then averaged to obtain the final SASV score, guiding the training of the parallel network.


\section{Baseline Systems}
\begin{figure*}[t]
\includegraphics[scale=0.45]{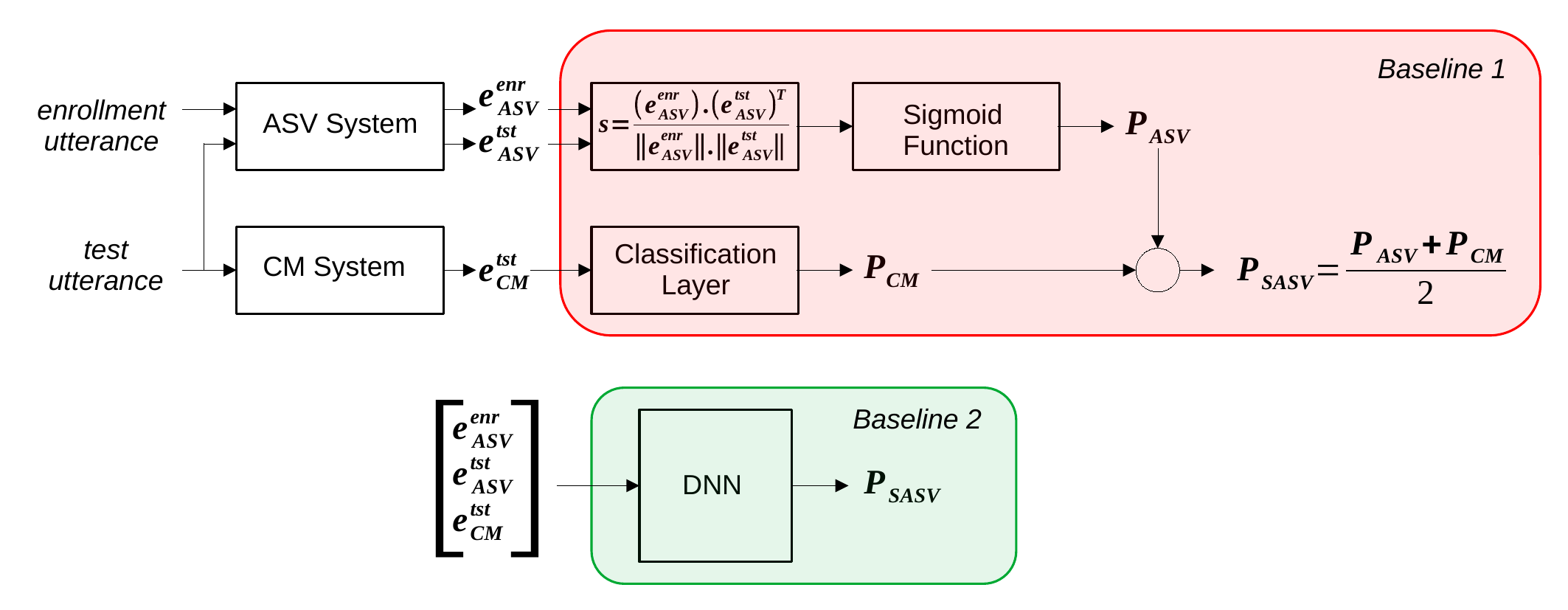}
\caption{Baseline Systems. \emph{Baseline 1} and \emph{Baseline 2} refer to score fusion and embedding fusion, respectively}\label{fig1:baseline}
\end{figure*}  
\subsection{Score Fusion}
Given a pair of enrollment and test utterances, $s^e$ and $s^t$, the SASV system aims to classify $s^t$ into $t_\mathrm{SASV}\in \{0,\: 1\}$ where $t_\mathrm{SASV}=1$ corresponds to the case where $s^e$ and $s^t$ belong to the same speaker and $s^t$ is bonafide, whereas $t_\mathrm{SASV}=0$ corresponds to the case where $s^e$ and $s^t$ originate from different speakers or $s^t$ is spoofed. Generally, two independent classifiers (one for the ASV task and another for the CM task) are developed to build an SASV system. Deep speaker embeddings, $e^{\mathrm{enr}}_\mathrm{ASV}$ and $e^{\mathrm{tst}}_\mathrm{ASV}$, are extracted using the ASV system from the enrollment ($s^e$) and test utterances ($s^t$), respectively. Then the cosine similarity is computed using these embeddings to represent the degree of similarity between enrollment and test embeddings. However, since cosine similarity takes values in the range $[-1, 1]$, it is processed through sigmoid function to fit its range into $[0,1]$ and treat it as a probabilistic ASV score:
\begin{equation}
\label{eq:p_asv}
    P_{\mathrm{ASV}} = P\left( t_\mathrm{ASV} = 1 | e^{\mathrm{enr}}_\mathrm{ASV}, e^{\mathrm{tst}}_\mathrm{ASV} \right) = \sigma \left( \frac{ (e^{\mathrm{enr}}_\mathrm{ASV}) ^\mathrm{T} \cdot (e^{\mathrm{tst}}_\mathrm{ASV}) }{ \| e^{\mathrm{enr}}_\mathrm{ASV} \| \cdot \| e^{\mathrm{tst}}_\mathrm{ASV} \|} \right) 
\end{equation}
where $\sigma$ is the sigmoid function and $t_\mathrm{ASV}\in \{0, 1\}$, with $t_\mathrm{ASV}=1$ when both enrollment and test embeddings belong to the same speaker and $t_\mathrm{ASV}=0$ otherwise.

For the CM task in turn, a CM embedding, $e_\mathrm{CM}^\mathrm{tst}$, is extracted from the CM system and it is classified into $t_\mathrm{CM}\in \{0, 1\}$, where $t_\mathrm{CM}=1$ indicates that the test utterance is bonafide and $t_\mathrm{CM}=0$ indicates that the test utterance is spoofed. This classification is performed using the CM score, which is computed in a probabilistic manner (e.g., the output of a softmax or sigmoid activation function) as follows:
\begin{equation}
\label{eq:p_cm}
    P_\mathrm{CM} = P(t_\mathrm{CM}=1|e_\mathrm{CM}^\mathrm{tst})
\end{equation}

From the SASV perspective, it is evident that $t_\mathrm{SASV}=1$ if and only if both $t_\mathrm{ASV}=1$ and $t_\mathrm{CM}=1$. Therefore, according to the law of total probability, the output probabilities of the two systems (ASV and CM systems) can be probabilistically fused to build an SASV system by simply averaging the output probabilities:
\begin{equation}
    P_\mathrm{SASV} = P(t_\mathrm{SASV}=1|e^{\mathrm{enr}}_\mathrm{ASV}, e^{\mathrm{tst}}_\mathrm{ASV},e^{\mathrm{tst}}_\mathrm{CM}) =  \frac{1}{2} \left( P_\mathrm{ASV} + P_\mathrm{CM} \right)  
\end{equation}
which is similar to the Baseline 1 score fusion method proposed for the SASV 2022 challenge as described in \cite{Shim2022} and summarized in Fig.~\ref{fig1:baseline}






\subsection{Embedding Fusion}


In the baseline embedding fusion approach, the embeddings obtained from the ASV ($e^{\mathrm{enr}}_\mathrm{ASV}$, $e^{\mathrm{tst}}_\mathrm{ASV}$) and CM ($e^{\mathrm{tst}}_\mathrm{CM}$) models are stacked together to form a single high-dimensional feature vector and then this feature vector is classified into $t_\mathrm{SASV}\in\{0,1\}$ using a simple DNN model consisting of several fully connected layers, as shown in Fig.~\ref{fig1:baseline}. This structure is similar to the Baseline 2 system proposed in the SASV2022 challenge and it processes the combined embeddings to produce SASV score, $P_\mathrm{SASV}$. These scores are then used to evaluate the system's performance, leveraging the combined strengths of the different model embeddings to improve accuracy and robustness.


\section{Proposed Method}
Although the embedding fusion method described in the previous section is a reasonable approach for the SASV task, a single DNN system may struggle to learn the most representative features for the SASV problem from the stacked ASV and CM embeddings due to its inherent architectural constraints. According to several preliminary experiments carried out, it is observed that different combinations of ASV and CM embeddings used for the embedding fusion yields different SASV performance. This corresponds to the fact that although the same DNN model is used to train the SASV system, it learns different level of SASV information according to input features. Therefore, we propose to use an SASV system based on the embedding fusion approach which consists of two parallel DNNs operating simultaneously and optimized jointly. 

\begin{figure}[h]
\includegraphics[scale=0.5]{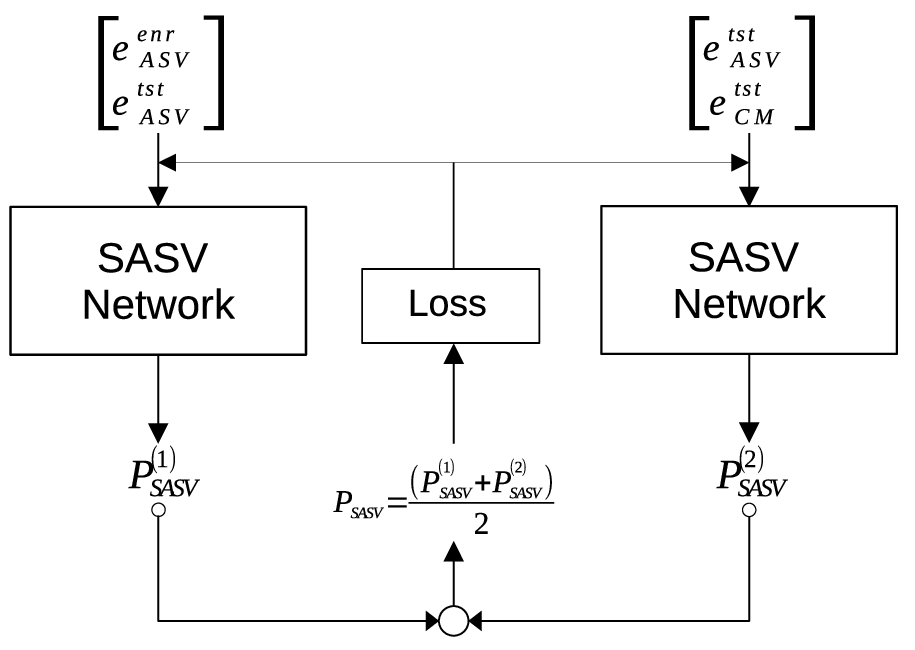}
\caption{Proposed Parallel Model}
\label{fig1:parallel_model}
\end{figure}

The system illustrated in Fig. \ref{fig1:parallel_model} demonstrates our proposed parallel model designed for SASV. The core concept behind this model is to utilize existing ASV and CM embeddings, which are fed into two identical neural networks running in parallel. Despite the identical structure of these networks—featuring the same number of layers and neurons—their inputs can differ, resulting in varying input dimensions. This setup allows the model to leverage diverse information sources while maintaining consistency in processing.

In this parallel architecture, concatenated embeddings are fed into each SASV network. Both networks process their respective inputs independently, producing SASV probabilities at their outputs. The outputs from these networks are represented as \( P^{(1)}_{\mathrm{SASV}} \) and \( P^{(2)}_{\mathrm{SASV}} \). To derive a final SASV probability, these individual probabilities are averaged:

\begin{equation}
 P_{\mathrm{SASV}} = \frac{P^{(1)}_{\mathrm{SASV}} + P^{(2)}_{\mathrm{SASV}}}{2}    
\end{equation}


This averaging process ensures that the final SASV probability \( P_{\mathrm{SASV}} \) reflects a balanced consideration of the outputs from both networks, potentially improving the robustness of the SASV system.

The model optimization is guided by the a-DCF loss function, as proposed in \cite{kurnaz2024optimizing} and BCE loss. The a-DCF loss function is defined as:
\begin{equation}
\label{adcf}
\begin{aligned}
\mathcal{L}_\mathrm{a\!-\!DCF}(\tau_\mathrm{sasv}) = & \; C_{\mathrm{miss}}^{\mathrm{tar}} \cdot \pi_{\mathrm{tar}} \cdot \hat{P}_{\mathrm{miss}}^{\mathrm{tar}}(\tau_\mathrm{sasv}) \\
& + C_{\mathrm{fa}}^{\mathrm{non}} \cdot \pi_{\mathrm{non}} \cdot \hat{P}_{\mathrm{fa}}^{\mathrm{non}}(\tau_\mathrm{sasv}) \\
& + C_{\mathrm{fa}}^{\mathrm{spf}} \cdot \pi_{\mathrm{spf}} \cdot \hat{P}_{\mathrm{fa}}^{\mathrm{spf}}(\tau_\mathrm{sasv}),
\end{aligned}
\end{equation}

\noindent where \( C_{\mathrm{miss}}^{\mathrm{tar}} \) is the cost of a miss, \( C_{\mathrm{fa}}^{\mathrm{non}} \) is the cost of nontarget false alarm and \( C_{\mathrm{fa}}^{\mathrm{spf}} \) is the respective costs for spoof false alarms. \( P_{\mathrm{miss}}^{\mathrm{tar}} \) is the miss rate, \( P_{\mathrm{fa}}^{\mathrm{non}} \) is nontarget false alarm rate and \( P_{\mathrm{fa}}^{\mathrm{spf}} \) is the false alarm for spoof trials. \( \pi_{\mathrm{tar}} \) is the prior probability of the target, \( \pi_{\mathrm{non}} \) is the prior probability of nontarget and \( \pi_{\mathrm{spf}} \) is the prior probability of spoofing. The miss rate ($\hat{P}_{\mathrm{miss}}^{\mathrm{tar}}$) and the false alarm rates ($\hat{P}_{\mathrm{fa}}^{\mathrm{non}}$, $\hat{P}_{\mathrm{fa}}^{\mathrm{spf}}$) are the \emph{softened} version of miss and false alarm rates as proposed in \cite{kurnaz2024optimizing}. $\tau_\mathrm{sasv}$ is the detection threshold for the SASV system.

Additionally, the Binary Cross-Entropy (BCE) loss function is employed to further refine the training process. The BCE loss is defined as:
\begin{equation}
\label{eq:bce_loss}
    \mathcal{L}_{\mathrm{BCE}} = - \frac{1}{N} \sum_{i=1}^{N} \left[ y_i \log(\hat{y}_i) + (1 - y_i) \log(1 - \hat{y}_i) \right]
\end{equation}

\noindent where \( N \) is the number of samples, \( y_i \) is the true label, and \( \hat{y}_i \) is the predicted probability.

To enhance the model's capability to handle the SASV task effectively, we employ the average of the a-DCF and BCE losses as our combined loss function:
\begin{equation}
\label{eq:combined_loss}
\mathcal{L}_{\mathrm{combined}} = \frac{1}{2} \bigg ( \mathcal{L}_\mathrm{a-DCF}(\tau_\mathrm{sasv}) + \mathcal{L}_\mathrm{BCE} \bigg ).
\end{equation}

This combined loss function aims to balance the trade-offs between the detection costs and the classification accuracy, leading to a more robust and efficient model optimization process.




\section{Experimental Setup}
\subsection{Dataset}

In the ASVspoof5 challenge closed condition, the participants are restricted to use ASVspoof5 data and Voxceleb2 data to train CM and ASV systems, respectively. In the open condition in turn, participants are allowed to use external data or pre-trained models. For the closed condition submission, we use ASVspoof5 data to train the CM system and Voxceleb2 data to train the ASV system. For the open condition submission, we use pre-trained WavLM model to extract ASV embeddings. 


\subsection{ASV Systems}
In the experiments we utilize two models to extract ASV embeddings: (i) ECAPA-TDNN \cite{Desplanques_2020} model trained using the Voxceleb2 dataset and (ii) pre-trained WavLM \cite{Chen2021WavLM} model. ECAPA-TDNN model is used to extract the $192$-dimensional ASV embeddings from the enrollment and test utterances. These embeddings are then used to either compute the ASV score using the Eq.~(\ref{eq:p_asv}) or to use the embedding fusion method or the proposed parallel SASV system.

The success of self-supervised speech foundation models has motivated us to utilize the WavLM architecture, as described in \cite{Chen2021WavLM}. WavLM is based on the HuBERT~\cite{hsu2021hubert} architecture, with a focus on modeling spoken content while preserving speaker identity. The training strategy for this architecture involves utterance mixing, including the generation of overlapped utterances, to enhance speaker discrimination. In this work, we use the pre-trained model\footnote{\url{https://huggingface.co/docs/transformers/en/model_doc/wavlm}} to extract $768$-dimensional feature representations. These features are obtained from the output of the last transformer layer of the WavLM model, sampled every $20$ milliseconds. The features for each utterance are then mean-pooled to produce utterance-level embeddings. These embeddings, obtained from both the enrollment and test sets, are subsequently used with baselines and the proposed framework to perform the SASV task.

\begin{table*}[h!]
    \centering
    \caption{Results of different systems and their configurations. The columns include the system identifier, method used (score fusion or embedding fusion), the ASV systems employed, the CM systems used, the optimization method applied, and the a-DCF results for both development and progress sets. The table compares various combinations of ECAPA-TDNN, WavLM, and AASIST models under different fusion strategies and optimization techniques. The smallest a-DCF value within each sub-group is bolded and globally smallest a-DCF values are underlined and bolded.}
    \vspace{0.1 cm}
    \begin{tabular}{|c|c|c|c|c|c|c|}
        \hline
        \multirow{2}{*}{\textbf{System}} & \multirow{2}{*}{\textbf{Method}} & \multirow{2}{*}{\textbf{ASV System}} & \multirow{2}{*}{\textbf{CM System}} & \multirow{2}{*}{\textbf{Loss}} & \multicolumn{2}{c|}{\textbf{a-DCF}} \\
        \cline{6-7}
        & & & & & \textbf{Dev} & \textbf{Prog.} \\
        \hline
        \hline
        S1 & Score Fusion & ECAPA-TDNN & AASIST & - & 0.2268 & 0.3767 \\
        \hline
        \hline
        S2 & Embedding Fusion & ECAPA-TDNN & AASIST & a-DCF & 0.2776 & 0.4046 \\
        \hline
        S3 & Embedding Fusion & ECAPA-TDNN & AASIST & BCE & 0.2830 & 0.3956 \\
        \hline
        S4 & Embedding Fusion & ECAPA-TDNN & AASIST & a-DCF + BCE & \textbf{0.2754} & \textbf{0.3937} \\
        \hline
        \hline
        S5 & Embedding Fusion & WavLM & AASIST & a-DCF & \textbf{0.2677} & 0.4084 \\
        \hline
        S6 & Embedding Fusion & WavLM & AASIST & BCE & 0.2951 & 0.3927 \\
        \hline
        S7 & Embedding Fusion & WavLM & AASIST & a-DCF + BCE & 0.2801 & \textbf{0.3831} \\
        \hline
        \hline
        S8 & Embedding Fusion & ECAPA-TDNN, WavLM & AASIST & a-DCF & 0.2316 & \textbf{0.3790} \\
        \hline
        S9 & Embedding Fusion & ECAPA-TDNN, WavLM & AASIST & BCE & 0.2204 & 0.4364 \\
        \hline
        S10 & Embedding Fusion & ECAPA-TDNN, WavLM & AASIST & a-DCF + BCE & \textbf{0.2088} & 0.3949 \\
        \hline
        \hline
        S11 & Proposed Parallel Model & ECAPA-TDNN & AASIST & a-DCF + BCE & \textbf{0.1692} & \textbf{0.2492} \\
        \hline
        S12 & Proposed Parallel Model & WavLM & AASIST & a-DCF + BCE & 0.1820 & 0.3253 \\
        \hline
        S13 & Proposed Parallel Model & ECAPA-TDNN, WavLM & AASIST & a-DCF + BCE & 0.2482 & 0.3582 \\
        \hline
        \hline
        S14 & Score Fusion (S11 + S12) & ECAPA-TDNN, WavLM & AASIST & - & \textbf{\underline{0.1250}} & \textbf{\underline{0.2129}} \\
        \hline
    \end{tabular}
    \label{tab:results_devprog}
\end{table*}

\subsection{CM System}
AASIST \cite{jung2022aasist} model is used as the CM system in the experiments and it is trained using the ASVspoof5 training subset. AASIST processes raw waveforms to learn high-dimensional spectro-temporal feature maps. It then extracts graph nodes from these feature maps in both temporal and frequency domains. By employing a stack node that assimilates information from all nodes, the final CM embedding is produced by concatenating the mean and maximum values of various nodes. We use AASIST model to extract the 
$160$-dimensional CM embeddings from the test utterances. These embeddings are then used in the baseline and proposed fusion framework to perform the SASV task.

\begin{figure*}
    \centering
    \includegraphics[scale=0.33]{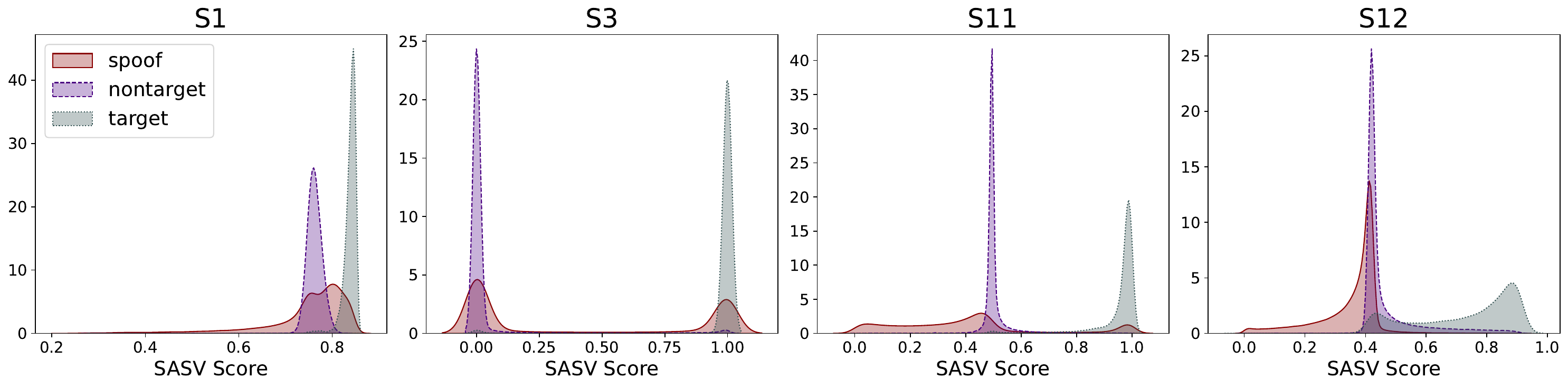}
    \caption{The graph illustrates the score distributions for the \emph{S1, S3, S11}, and \emph{S12} systems based on class labels. \emph{S1} shows the distribution of the averaged scores obtained from the ECAPA-TDNN and AASIST models for ASV and CM. \emph{S3} depicts the distribution of prediction scores resulting from a DNN model trained with BCE loss, using embeddings extracted from the ECAPA-TDNN and AASIST models as input. \emph{S11} presents the distribution of prediction scores from our proposed parallel DNN model, which was trained using a combination of a-DCF and BCE loss functions, utilizing embeddings from the ECAPA-TDNN and AASIST models. \emph{S12} illustrates the distribution of prediction scores from our proposed parallel DNN model, trained with a combination of a-DCF and BCE loss functions, using embeddings obtained from the WavLM and AASIST models.}
    \label{fig:score_distributions}
\end{figure*}

\subsection{SASV System}

In the Baseline system as shown in Fig. \ref{fig1:baseline}, embedding fusion was performed, and these embeddings were used as inputs to a DNN structure. This DNN structure consists of an input layer, three hidden layers, and an output layer. The hidden layers contain $256$, $128$, and $64$ nodes, respectively, while the output layer contains a single node.

In the parallel structure as shown in Fig. \ref{fig1:parallel_model}, embeddings were used as inputs to two parallel DNN structures. These structures were trained independently. After the training phase, the SASV scores obtained from the outputs of the two DNNs were averaged. This average score was then used to optimize both parallel DNN structures. All parallel models are almost identical except the dimension of the inputs. The hidden layers contain $256$, $128$, and $64$ nodes, respectively, while the output layer contains a single node. All other details are same as in baseline.

\section{Results and Discussion}
\subsection{Results on Development and Progress Sets}

Table \ref{tab:results_devprog} 
 presents the minimum a-DCF values achieved with various system configurations on the development and progress sets. In our experiments, we initially employed the baseline score fusion and embedding fusion approaches for the SASV task. Consistent with the findings reported for the SASV 2022 challenge \cite{Shim2022} using the SASV-EER metric, score averaging outperforms embedding fusion in terms of the minimum a-DCF metric for both the development and progress sets. For instance, score fusion achieves a minimum a-DCF value of $0.3767$ on the progress set, whereas embedding fusion results in an a-DCF value of $0.3956$.

Next, we examine the impact of different loss functions on optimizing the network parameters during the training of the embedding fusion network.  Specifically, we replace the conventional BCE loss with the a-DCF loss defined in  Eq.~(\ref{adcf}) as proposed in \cite{kurnaz2024optimizing}  (S2 model in the table) and combined loss function defined in Eq.~(\ref{eq:combined_loss}) (S4 model in the table). The results show that training the system with combination of traditional BCE loss and a-DCF loss (achieving a minimum a-DCF value of 0.3937 on the progress set) outperforms both the standalone BCE loss (a-DCF value of $0.3956$) and the standalone a-DCF loss (a-DCF value of $0.4046$). 

In the experiments, we further investigate the effect of using the pre-trained WavLM model to extract ASV embeddings (systems S5-S7 in Table~\ref{tab:results_devprog}). The results in the table indicate that although the WavLM model was not initially trained for speaker recognition, its embeddings still capture speaker-specific information. For instance, using ASV embeddings extracted from the ECAPA-TDNN model with CM embeddings from the AASIST system yields a minimum a-DCF value of $0.3956$ with the BCE loss. However, when the embeddings from the WavLM model are used instead of those from ECAPA-TDNN, the a-DCF value decreases to 0.3927. This demonstrates that the WavLM model's embeddings effectively preserve speaker information. Similar to the baseline systems (S2-S4 in the table) that utilize ECAPA-TDNN for the ASV system, combining BCE and a-DCF losses further enhances SASV performance for the WavLM model.

Noting the interesting results achieved with the WavLM model as the ASV system, we further investigate whether the embeddings from WavLM and ECAPA-TDNN provide complementary speaker information. To explore this, we combined ASV embeddings from both models (ECAPA-TDNN and WavLM) with CM embeddings from the AASIST system and trained the baseline embedding fusion network using various loss functions (Systems S8-S10 in the table). Regardless of the loss function used, the combination of both embeddings significantly reduces a-DCF values compared to using single ASV systems (either ECAPA or WavLM alone) on the development set. For example, with the BCE loss function, the a-DCF values are $0.2830$ for ECAPA embeddings and $0.2951$ for WavLM embeddings. However, combining the embeddings from both systems achieves an a-DCF value of $0.2204$, representing an approximate $22\%$ relative improvement over the ECAPA-TDNN embeddings. This demonstrates that the embeddings from WavLM and ECAPA-TDNN capture complementary speaker-specific information.

Finally, in our experiments on the development and progress sets, we apply the proposed parallel model for the SASV task using various ASV embedding extractors. We use a combination of BCE and a-DCF losses in the parallel model, as it generally provides better performance compared to standalone losses based on previous experiments. The proposed parallel model, consisting of three layers, significantly outperforms the baseline systems. However, in our S11 implementation, we used a two-layer architecture with $128$ and $64$ nodes, as it yielded better experimental results. The best results with this method are achieved using ECAPA-TDNN as the ASV embedding extractor (S11 in the table) on both the development and progress sets. This demonstrates the effectiveness of the proposed approach and indicates that each branch of the parallel network captures different aspects of SASV information. The last row of the table shows the score fusion of the two best-performing systems, both of which are based on the proposed parallel model with different ASV embeddings. Combining the scores of these systems using simple score summation results in approximately a $14\%$ relative improvement in SASV performance over the individual proposed method.

To further investigate the performance of the various systems used in this study, Fig.~\ref{fig:score_distributions} illustrates the score distributions for the baseline systems (both score and embedding fusion methods) as well as the proposed parallel model. The figure clearly shows that the proposed method achieves better separation of target, nontarget, and spoof scores compared to the baseline methods, highlighting the effectiveness of the proposed technique.

\subsection{Results on Evaluation Set}
Based on extensive experiments conducted on the development and progress subsets, we evaluated the SASV scores on the evaluation subset using the proposed parallel model with embeddings from the ECAPA-TDNN and AASIST models for the closed condition submission. For the open condition submission, we employed score fusion of the scores from parallel models S11 and S12. The a-DCF values for our submissions on the evaluation set are summarized in  Table~\ref{tab:results_eval}.

\begin{table}[h!]
\centering
\caption{The evaluation set results of the \emph{S11} and \emph{S14} systems, which provided the top two results based on the progress outcomes from different configurations. \emph{S11} and \emph{S14} systems are used as the \emph{Close Condition} model and \emph{Open Condition} model, respectively, for the SASV (Track-2) subtask of the ASVspoof5 challenge.}
\vspace{0.2cm}
\begin{tabular}{|c|c|c|}
    \hline
    \textbf{System} & \textbf{Method} & \textbf{Eval a-DCF} \\
    \hline
    S11 & Proposed Parallel Model & 0.5130 \\
    \hline
    S14 & Score fusion (S11 + S12)  & 0.4581 \\
    \hline
\end{tabular}
\label{tab:results_eval}
\end{table}

Our submission using the parallel model with ECAPA-TDNN and AASIST embeddings (S11) achieved an a-DCF score of $0.5130$. In contrast, score-level fusion demonstrated improved performance, yielding an a-DCF of $0.4581$. This trend was consistent across all partitions, supporting the conclusion that we have developed a robust and effective method for spoofing-aware speaker verification.

\section{Conclusion}
In this paper, we presented the parallel DNN model designed for the SASV task in the ASVspoof5 challenge. Our proposed method demonstrated significant performance enhancements compared to traditional score-sum and embedding fusion approaches. Additionally, we showed that despite not being originally trained for speaker recognition, the embeddings from the WavLM model effectively capture speaker-specific information, contributing to improved SASV performance.

\section{Acknowledgement}
The work of Jagabandhu Mishra was supported by the Academy of Finland (Decision No. 349605, project "SPEECHFAKES").

\bibliographystyle{IEEEtran}
\bibliography{bibliography}

\end{document}